\documentclass{icrc}

\usepackage{times}
\usepackage{graphicx} 

\begin{document}

\title{Limits on Pulsar Parameters for Pulsed detections with H.E.S.S.}
\author[1]{O.C. de Jager}
\affil[1]{Unit for Space Physics, Potchefstroom University, Potchefstroom,
2520, South Africa}
\author[2]{A. Konopelko}
\affil[2]{Max Planck Institut f\"ur Kernphysik, Postfach 103980, D-69029
Heidelberg, Germany}
\author[1]{B.C. Raubenheimer}
\author[1]{B. Visser}

\correspondence{O.C. de Jager: okkie@fskocdj.puk.ac.za}

\firstpage{1}
\pubyear{2001}


\maketitle

\begin{abstract}
The non-detection of pulsed sub-TeV $\gamma$-rays from EGRET pulsars
proves that the EGRET pulsed spectra of all $\gamma$-ray pulsars
should terminate at energies below a few hundred GeV. The spectrum 
of a typical integrated pulse profile predicted by the polar cap model 
resemble typically a hard component, followed by a super exponential 
cutoff between 1 MeV (PSR B1509-58) and tens of GeV (e.g. Crab, 
PSR B1951+32 etc). Outergap models predict a hard low flux component 
extending to TeV energies, and the stereoscopic property of the
H.E.S.S. (High Energy Stereoscopic System) ground-based detector
(under construction) would have the advantage to discriminate against 
the background above 50-100 GeV, so that such a second component may
be detectable. However, the challenge posed for any ground-based 
$\gamma$-ray detection is to prove that the instrument can detect a 
pure polar cap origin, whereas an outergap mechanism would provide 
little challenge given the rapid increase in the effective area
$A(E)$ with increasing energy $E$ for Cerenkov telescopes.

Using a topological trigger in the non-imaging mode, we show that 
H.E.S.S. should be able to detect pulsed emission from PSR B1706-44
within a few hours if the cutoff energy is above 30 GeV as suggested 
by EGRET observations. The recently detected radio pulsar PSR
J1837-0604 (pulsar period: 96 ms) associated with the unidentified 
EGRET source GeV J1837-06010 should also be detectable within a few 
hours if the source is pulsed and if its cutoff is similar to that of 
PSR B1706-44. H.E.S.S. should even be able to image middle-aged, 
low-multiplicity pulsars for which the mean photon energy is expected 
to be well above 10 GeV. Such observations should provide important
constraints on the final evolutionary status of $\gamma$-ray pulsars 
and millisecond pulsars in general. 
\end{abstract}

\section{Introduction}
Basic pulsar electrodynamics involves a definition of the open field
lines, which close outside the light cylinder, so that a polar cap 
potential drop $\Delta V \propto B_s/P^2$ is in excess of $10^{12}$ eV, 
where $B_s$ is the surface field strength and $P$ the pulsar period. 
The Goldreich Julian pulsar current from the polar cap scales 
similarly as $I_{\rm GJ} \propto B_s/P^2$), and is responsible for the 
extraction of electrons and possibly Fe ions from the surface of the 
polar cap, where a component of the electric field parallel to 
${\bf B}$ exists. The product of $\Delta V$ and $I_{\rm GJ}$ is then
comparable to the magnetic dipole spindown power.

It is therefore natural to follow the curvature $\gamma$-ray emission
from accelerated electrons above the polar cap, resulting in a cone-like
beam of $\gamma$-ray emission above the polar cap until
magnetic pair production no longer dominates the pair production process.
Daugherty \& Harding (1996) were able to reproduce the observed
pulse profiles and spectra for such processes, but a
natural consequence of the polar cap process is a super exponential
cutoff of the spectrum as discussed by Nel \& de Jager (1995). 

In competition with the polar cap model, is the so-called
outergap model, whereby a potential drop may develop around 
the ${\bf \Omega \cdot B}=0$ ``null surface'', which is at a much larger
distance from the pulsar compared to the polar cap emission region.
Spectra can be self-consistently derived for outer gaps
(Hirotani 2001), but it remains to be shown how the outergap geometry
can be unified with the electrodynamics. In this case we find that
photon-photon pair production and the available acceleration potential
determines the cutoff. Both a synchrotron and a much higher 
energy inverse Compton component can then escape from the outergap
to the observer. 

A generic model (polar cap and/or outergap) for the 
tails of pulsed differential spectra is then given by
\begin{eqnarray*} 
\frac{dN}{dE} & = & K_1E^{-\Gamma_1}\exp(-(E/E_1)^b)\\
& & \mbox{}+K_2E^{-\Gamma_2}\exp{(-(E/E_2)^c.)}
\end{eqnarray*}
The second component would be absent in the case of pure polar
cap $\gamma$-ray emitters, with the additional signature of 
a super exponential cutoff ($b\geq 1$ and $K_2=0$). An outergap
origin can be interpreted in terms of a non-zero $K_2$, but with
slower rollovers compared to polar cap models, since the outerap
absorption process is controlled by photon-photon pair production,
which has a weaker energy dependence compared to magnetic pair 
production.

Timing and spectroscopic observations in the 10 GeV to 10 TeV region 
can therefore discriminate between such models, and GLAST is expected 
to play a key role in this regard, but next generation telescopes such 
as H.E.S.S., MAGIC and CELESTE, STACEE (at their full capacity) should 
be able to steal some of the limelight in the meantime, with H.E.S.S. 
and CANGAROO IV, VERITAS in the best position to test outergap models 
above 50-100 GeV, given their stereo capability for efficient
background rejection.

\section{Gamma-Ray Pulsar Parameters}
To obtain conservative estimates for the detection rate, we have
to employ the most conservative model. We therefore assume
that the polar cap mechanism is responsible for the pulsed $\gamma$-ray
emission, so that our generic model reduces to a single component: 
\begin{equation}
dN_{\gamma}/dE=K(E/E_n)^{-g}exp(-(E/E_o)^b). \noindent
\end{equation}
Whereas pulsar photon spectral indices between $g=1.4$ and 2.1
are observed, harder spectra are theoretically possible for middle-aged
pulsars (A.K. Harding, 2000, personal communication to O.C. de Jager).
The constant $K$ represents the monochromatic flux at the normalising
energy $E_n\ll E_o$. We will normalise spectra at $E_n$ near 1 GeV. 

It was shown that the total $\gamma$-ray pulsed luminosity
$L_{\gamma}$ (pulsed) scales roughly with the Goldreigh Julian pulsar 
current $B_s/P^2$, which, in turn, is proportional
to $\dot{E}^{1/2}$, with $\dot{E}$ the spindown power. Neglecting
the differences in the spectral index and cutoff energies, we even find
that the normalization constant $K$ (in units of
cm$^{-2}$s$^{-1}$GeV$^{-1}$) at 1 GeV scales with basic pulsar parameters:
Fitting $K$ as a power law function of the spindown power, we find  
(normalised to the EGRET pulsars):
\begin{equation}
K=10^{-17.8}(\dot{E})^{0.305}(d_{\rm kpc})^{-2}
\end{equation}
The spindown power is in units of ergs/s. It is interesting to notice that
Vela and PSR B1706-44 (with similar values for $\dot{E}$) give the
largest scatter as a result of their significant differences in flux
and distance. In fact, beaming and line-of-sight effects contribute to
the large scatter.

Whereas the total $\gamma$-ray pulsed flux scales with pulsar current
$I_{\rm GJ}$ ($\propto \dot{E}^{1/2}$), we therefore find that the GeV
flux (or luminosity), to a first order, scales as
\begin{equation}
L_{\gamma}(GeV)\propto (I_{\rm GJ})^{0.6}
\end{equation}
This finding is in principle consistent with the claim of Nel \& de
Jager (1995) that the mean photon energy increases with increasing
pulsar age. Thus, older (lower $\dot{E}$) pulsars become relatively 
brighter in the GeV region, which explains the weaker dependence
of the GeV flux on current (or spindown power). It is thus
clear that ground-based $\gamma$-ray observations may have the best chance
of detecting middle aged pulsars, provided that the beaming is favorable,
the distance is not too large, and the cutoff energy is well above 10 GeV.
All these constraints are reasonably within reach for future
ground-based $\gamma$-ray observations. 

\begin{table*}[t]
\begin{center}
\caption{Gamma-ray spectral parameters above 1 GeV and corresponding
H.E.S.S. rates and observation time for detection. Spectral references
from Macomb \& Gehrels (1999).}
\begin{tabular}{lccccccc}
\hline
Object & $k$ ($\times 10^{-8}$) & $g$ & $E_o$ & $b$ & $F(>1\;{\rm GeV})$ 
& $R_p$ & $T$ (10-hour \\
&(cm$^{-2}$s$^{-1}$GeV$^{-1}$) & & (GeV) & & (cm$^{-2}$s$^{-1}$) 
& (hour$^{-1}$) & days)\\
\hline
Crab         & 24.0   & 2.08 & 30  & 2   & 22   & 100  & 3 \\
Vela         & 138    & 1.62 & 8.0 & 1.7 & 148  & 8    & 400 \\
Geminga      & 73.0   & 1.42 & 5.0 & 2.2 & 76   & $\ll 1$& - \\
PSR B1951+32 & 3.80   & 1.74 & 40  & 2   & 4.9  & 180  & 1 \\
PSR B1055-52 & 4.00   & 1.80 & 20  & 2   & 4.5  & 8    & 420 \\
PSR B1706-44 & 20.5   & 2.10 & 40  & 2   & 20   & 240  & 1 \\
PSR J2229+61 & 4.8    & 2.24 & 40  & 2   & 3.9  & 32   & 25 \\ 
PSR J1420-60 & 6.9    & 2.02 & 40  & 2   & 6.9  & 110  & 2 \\
PSR J1837-06 & 5.5    & 1.82 & 40  & 2   & 6.7  & 190  & 1 \\
\hline
\end{tabular} 
\end{center} 
\end{table*}

\section{Detection capability of H.E.S.S. for pulsars}
De Jager et al. (2001) discussed the capability of H.E.S.S. (Hofmann et 
al (2001)) to detect pulsed emission from the EGRET pulsars, given the 
assumption of super exponential cutoffs as expected from polar cap 
emission. A detection within a single night (in general) restricts 
the number of independent frequencies to be searched, which enables
the identification of a single unique frequency, or, at least a number of
candidate frequencies which can be confirmed within a few days of
follow up observations. We will copy the list of de Jager et
al. (2001), but add the recent discoveries of radio pulsars associated 
with unidentified GeV EGRET sources.

Using the H.E.S.S. collection area vs. energy $A(E)$ for any 2-telescope
triggers (Konopelko 2000), we were able to calculate the expected
rates $R_p$ for pulsed $\gamma$-rays by integrating the product of 
$A(E)dN_{\gamma}/dE$ over all energies. The results for the six EGRET 
pulsars are also shown in Table 1 (indicated by ``$R_p$''). It is
clear that the rate for PSR B1706-44 is the largest of all pulsars 
if $E_o$ is not smaller than 40 GeV.
     
It was shown by de Jager, Swanepoel \& Raubenheimer (1989) and
de Jager (1994) that the basic scaling parameter for any test for
uniformity on the circle (given a test period) is given by
$x=p\sqrt{n}$, where $p=R_p/(R_b+R_p)$ is the pulsed fraction, with
$R_p$ the pulsed rate and $R_b$ the background rate. The total number of
events is given by $N=(R_p+R_b)T$, with
$T$ the observation time. In this case the test statistic
for uniformity for the general Beran (1969) class of tests is given by
$B=x^2\Phi_B +c$, where $\Phi_B$ is derived from the intrinsic pulse profile,
and c is the noise term. It was shown by Thompson (2001)
that the pulse profiles above 5 GeV consist mostly of two
narrow peaks, but given the spectral differences between the two peaks,
we will assume that only a single peak survives at the highest energies,
so that $\Phi_B=5.8$ if we assume a 5\% duty cycle, and
$B=Z^2_m$ test statistic with $m=10$ harmonics (see e.g. de Jager,
Swanepoel \& Raubenheimer 1989). In this case $c=20$.

A value of $x=3$ would introduce a $\sim 3\sigma$ DC excess in a 
spatial analysis, but assuming that we have no imaging capability
for $E_o$ near the detection threshold, we have to rely on a timing
analysis, which would give $Z^2_{10}\sim 73$, or a chance
probability of $7\times 10^{-8}$ if the period is known, but 0.03
after multiplying with the number of trials for a 6 hour observation
if searching for periods as short as 50 ms. A confirming run
(e.g. on a second night) 
should always be made to see if one of the few most 
significant periods from the previous run have repeated 
itself - in this case at the $\sim 10^{-7}$ level.

Using an additional topological software trigger, recently tested with 
the HEGRA system of 5 imaging atmospheric Cherenkov telescopes
(Lucarelli et al. (2001)) and selecting events by image size and
angular shape, we were able to reject $\sim 99.2\%$ of the 
triggered background events, while retaining 95\% of the 
source events below 50 GeV. From a total background rate of about 
1 kHz (Konopelko 2000), we get $R_b=8$ Hz. This allows us to calculate 
detection sensitivities for periodicities.

\begin{figure}
\includegraphics[width=8.3cm]{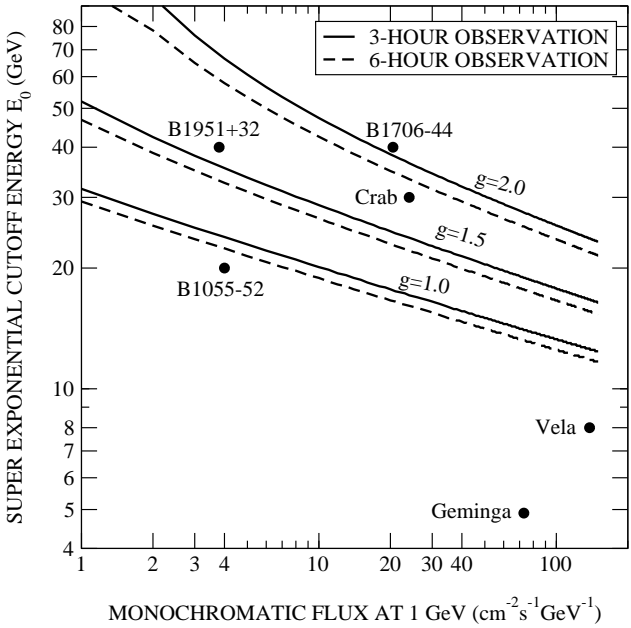} 
\caption{Parameter space 
($E_o$ vs $K$) for the detection of unknown pulsars within one night 
with H.E.S.S. using a timing analysis approach, and assuming 
a DC excess of $x=3$. The three
curves represent (from bottom to top) photon spectral indices of 1, 1.5
and 2.0. The solid line is for 3 hours of continuous observation, whereas
the dashed lines (for the same set of spectral indices) represent a six-hour
run.}
\end{figure}

From the GeV source catalogue, we find that
the fluxes of the galactic unidentified EGRET source 
(some may be pulsars - Lamb \& Macomb 1997) range from 
$F(>1\;{\rm GeV})$ = 1 to 25 $\times 10^{-8}$ cm$^{-2}$s$^{-1}$.
Figure 1 give the H.E.S.S. sensitivity for a wide
range of possible pulsar photon spectral indices 
between 1 and 2, and requiring a marginal detection within $T=3$ to
6 hours (assuming a minimum ``DC significance'' of $x=3$).
This figure also shows $E_o$ vs $K$, with the
latter within the EGRET range as discussed above.
Table 1 also shows $T$ calculated in the same way (in 10-hour shifts), but 
assuming the spectral parameters of individual pulsars.
It is clear the H.E.S.S. will only be able to detect pulsed emission
if $E_0$ exceeds 30 GeV, which is realized at least for PSR B1706-44
in the Southern Hemisphere.

\section{Three Young Pulsars Assocated with GeV Sources}
Halpern et al. (2001) reported on the discovery of an energetic
pulsar PSR J2229+6114 (51.6 s), which appears to be associated with the
GeV unidentified 
EGRET source GEV J2227+6101. D'Amico et al. (2001) also identified
two pulsars associated with another two
GeV EGRET sources: These are
PSR J1420-6048 (68 ms) associated with GeV J1417-6100, and
PSR J1837-0604 (96 ms) associated with GeV J1837-06010. Assuming
that the excess EGRET photons from these GeV sources are pulsed,
we have calculated the H.E.S.S. sensitivities for pulsed detections
assuming a cutoff of 40 GeV as shown in Table 1.
Since we do not know 
what the cutoff energy is, we also calculated the required observation
time as a function of the cutoff energy for these two pulsars
as shown in Figure 2. 

\begin{figure}[t]
\includegraphics[width=8.3cm]{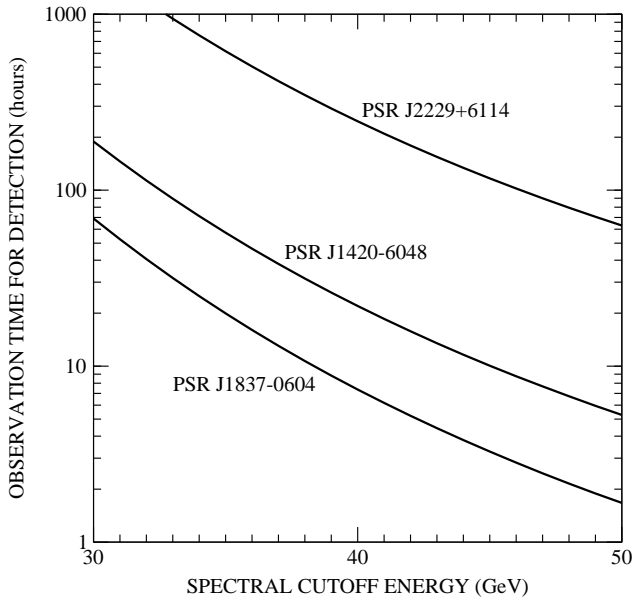} 
\caption{The observation time required for a pulsed detection 
by H.E.S.S. as a function of the spectral cutoff energy $E_0$ for
the three new radio pulsars associated with unidentified EGRET GeV
sources. The $\gamma$-ray spectrum associated with each pulsar is
given in Table 1. In each case it
was assumed that the EGRET excess is 100\% pulsed.}  
\end{figure}

The best candidate is PSR J1837-0604, and even for this pulsar
we require a cutoff energy similar to that of PSR B1706-44 (as large
as 30 - 40 GeV) if the pulsar is to be detected within one night
of observations.

\section{Conclusions}
By operating H.E.S.S. in a non-imaging 
topological trigger mode, we can accept low energy $\gamma$-rays, while
still rejecting $>99\%$ of the background, which allows us to detect
pulsations from pulsars if the spectral cutoff energy is above 30 GeV.
This, added with the hardness of the spectra below the cutoff, makes 
a detection within one night possible for a few pulsars.
We find that PSR B1706-44 should be the best candidate for 
the H.E.S.S. site in Namibia.

The detection of radio pulsars associated with unidentified EGRET sources
improves our confidence that many hard-spectrum galactic EGRET sources
may be due to pulsed emission from pulsars. We have discussed three recently
detected radio pulsars, and we find that PSR J1837-0604 (being a hard
GeV source) may also be detectable within a single night if the EGRET source
is pulsed and if the cutoff is above 30 - 40 GeV. In fact, one may even be
able to derive the period from a search in frequency if simultaneous
radio observations are not available. It can be shown that a slight difference
in cutoff energy can make an enormous difference in detection sensitivity:
increasing the cutoff energy by a relatively small amount can change the
situation from being a slave to radio information to that of determining
radio parameters independent of radio information.

Finally, there is quite a large population of millisecond pulsars, and some
of them are predicted to be $\gamma$-ray bright. Fierro 
et al. (1995) searched
for such pulsars in the EGRET data base, but found non despite the availability
of contemporary rotational parameters. Cascading above the polar cap is still
required to produce the radio emission, and some of them show pulsed 
magnetospheric X-ray emission, which is indicative of accelerating potentials
above the polar cap. One possibility is that the low magnetic fields associated
with millisecond pulsars results in a relatively small cascading multiplicity,
so that the mean photon energy which escapes from the polar cap is between
10 and 100 GeV - much closer to the primary curvature $\gamma$-ray energy
compared to canonical high-B pulsars. H.E.S.S. should be able to image such
sources above 50 GeV using its full stereo capability, and provide significant
upper limits if a source is not seen.

%

%




\begin{acknowledgements}
The South African group acknowledges the financial contribution
of the South African Department of Arts Culture and Science 
and Technology towards the H.E.S.S. project.

\end{acknowledgements}




\end{document}